\documentclass[conference]{IEEEtran}
\IEEEoverridecommandlockouts

\usepackage{cite}
\usepackage{amsmath,amssymb,amsfonts}
\usepackage{algorithmic}
\usepackage{graphicx}
\usepackage{textcomp}
\usepackage{xcolor}
\usepackage{multirow}
\usepackage{threeparttable}

\title{
Privacy Protected Contactless Cardio-respiratory Monitoring using Defocused Cameras during Sleep
}

\author{Yingen Zhu$^{1}$, Jia Huang$^{2}$, Hongzhou Lu$^{2,*}$, Wenjin Wang$^{1,*}$
\thanks{*This work is supported by the National Key R\&D Program of China (2022YFC2407800), General Program of National Natural Science Foundation of China (62271241), Guangdong Basic and Applied Basic Research Foundation (2023A1515012983), Shenzhen Science and Technology Program (JSGGKQTD20221103174704003), and Shenzhen Fundamental Research Program (JCYJ20220530112601003).}
\thanks{$^{1}$Department of Biomedical Engineering, Southern University of Science and Technology, China.}%
\thanks{$^{2}$Intensive Care Unit, The Third People's Hospital of Shenzhen, China.}%
\thanks{*Corresponding author: Hongzhou Lu (luhongzhou@fudan.edu.cn), Wenjin Wang (wangwj3@sustech.edu.cn)}
} 

\begin{document}

\maketitle
\thispagestyle{empty}
\pagestyle{empty}

\begin{abstract}

The monitoring of vital signs such as heart rate (HR) and respiratory rate (RR) during sleep is important for the assessment of sleep quality and detection of sleep disorders.
Camera-based HR and RR monitoring gained popularity in sleep monitoring in recent years.
However, they are all facing with serious privacy issues when using a video camera in the sleeping scenario. In this paper, we propose to use the defocused camera to measure vital signs from optically blurred images, which can fundamentally eliminate the privacy invasion as face is difficult to be identified in obtained blurry images.
A spatial-redundant framework involving living-skin detection is used to extract HR and RR from the defocused camera in NIR, and a motion metric is designed to exclude outliers caused by body motions.
In the benchmark, the overall Mean Absolute Error (MAE) for HR measurement is 4.4 bpm, for RR measurement is 5.9 bpm. Both have quality drops as compared to the measurement using a focused camera, but the degradation in HR is much less, i.e. HR measurement has strong correlation with the reference ($R \geq 0.90$). Preliminary experiments suggest that it is feasible to use a defocused camera for cardio-respiratory monitoring while protecting the privacy. Further improvement is needed for robust RR measurement, such as by PPG-modulation based RR extraction.

\end{abstract}




\section{Introduction}
Knowing the sleep quality is important for maintaining overall health and improving the functioning of mind and body\cite{lee2022relationship}. 
Insufficient and irregular sleep patterns may lead to stress, exacerbating chronic symptoms such as cardiovascular diseases\cite{powell2021obesity}, sleep apnea\cite{gu2019wifi}, and asthma\cite{braun2012bridging}. 
Monitoring vital signs during sleep like respiratory rate (RR) and heart rate (HR) plays an important role in diagnosing and treating the aforementioned sleep disorders\cite{thomas2014relationship}. Besides, it is valuable for preventing and diagnosing adult diseases like obesity, arrhythmias, and coronary artery diseases through continuous monitoring\cite{zhu2006real}. 
For high-risk patients undergoing inpatient surgeries, it can reduce the incidence of adverse postoperative outcomes\cite{tian2022impact}.
In home-based care settings, it is effective for monitoring babies at risk of sudden death due to periodic hypoxia and seniors at risk of cardio-cerebral infarction due to pre-existing cardiopulmonary diseases\cite{freed2017history}. 
Thus, fully-automatic continuous monitoring of cardio-respiratory activities has important application values in sleep-related fields.

Clinically, cardio-respiratory parameters measured by Polysomnography and contact biomedical sensors are commonly regarded as gold standards for sleep monitoring. However, these methods may cause discomfort for patients and inconvenience for physicians as the cumbersome sensors (e.g. electroencephalogram (EEG), electrocardiography (EOG), electromyography (EMG), electrocardiograph(ECG), etc.) need to be placed on patient's body. Moreover, the high cost of devices and complexity of operation make them impractical for home monitoring.
Recently, alternative contactless sleep monitoring technologies emerged, including thermal imaging\cite{hu2018combination}, radar\cite{turppa2020vital}, WIFI\cite{gu2019wifi}, etc. 
Among these, camera-based approach has gained much attention for sleep monitoring due to its low-cost and ubiquitous properties.
Besides, visual data provides rich semantic information, such as sleep posture, movements, and bed-exiting/falling, that can enhance sleep monitoring and detection of sleep-related disorders\cite{zhu2019vision}. 

\begin{figure}[t!]
        \centerline{\includegraphics[width=0.48\textwidth]
                {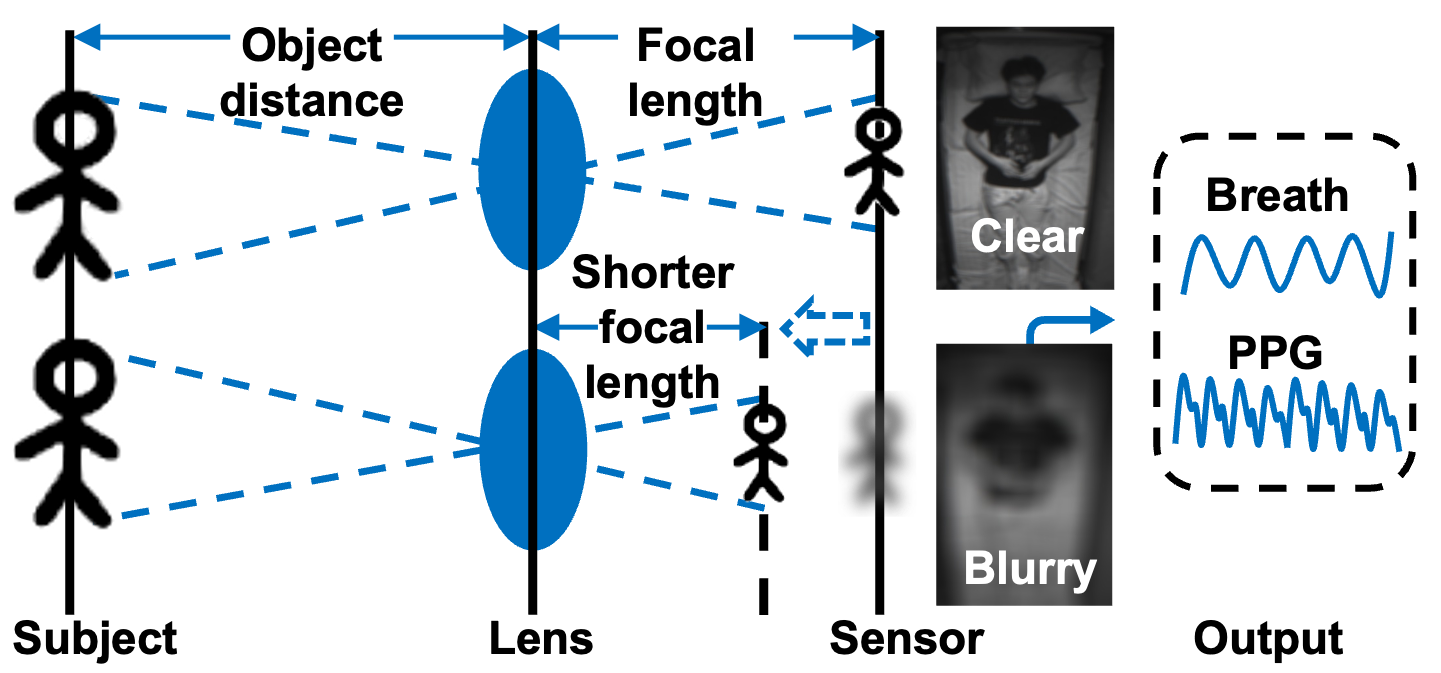}}
        \caption{First row: the imaging principle of a focused camera. 
        Second row: the principle of how the camera is defocused to obtain blurry images. }
        \label{fig:defocus}
        \vspace{-0.5cm}
\end{figure}

\begin{figure*}[t!]
        \centerline{\includegraphics[width=1\textwidth]
                {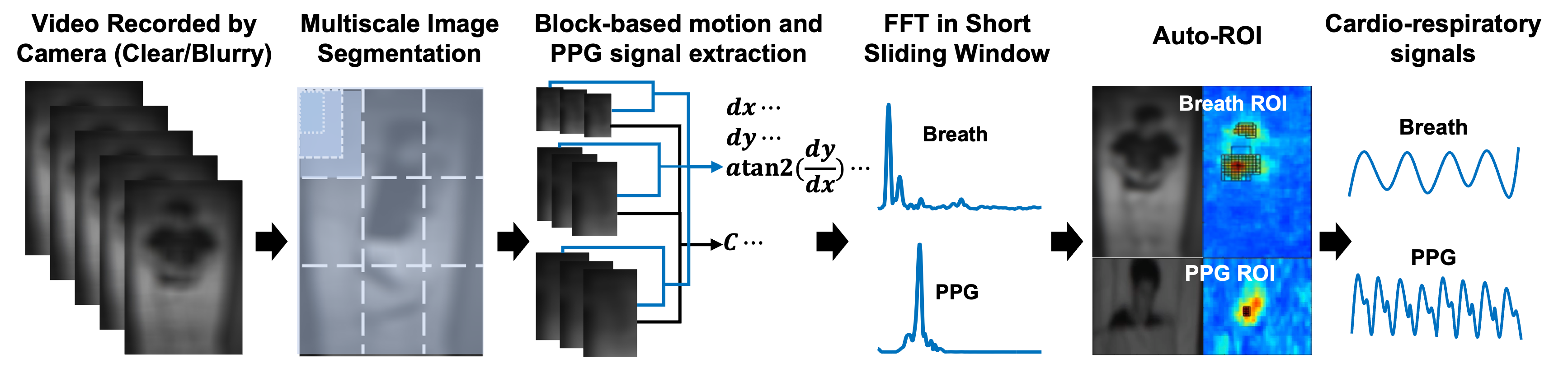}}
        \caption{The video processing framework for cardio-respiratory monitoring during sleep, designed for vital signs extraction in both focused (clear) and de-focused (blurry) conditions. \textbf{\emph{dx}} and \textbf{\emph{dy}} denote motion signals on the horizontal and vertical directions, and \textbf{\emph{C}} denotes the PPG signal.}
        \label{fig:method}
        \vspace{-0.4cm}
\end{figure*}

For camera-based monitoring, vital signs are usually extracted using the rPPG (remote photoplethysmography) method, an optical method that detects small changes in skin color induced by blood pulsation during the
cardiac cycle\cite{wang2015novel}. And the facial skin area is a common spot for vital signs measurement\cite{8575340}. 
However, it raises critical privacy issues as it associates a person's health condition with its identity, especially in privacy-sensitive scenarios like home-based sleep monitoring\cite{sunPrivacyPhysFacialVideoBased2022a}.
Earlier study proposed to eliminate the privacy issue using a single-element camera\cite{wang2018single}, but it sacrifices important spatial information required for semantic measurement, i.e. single-element approach can only measure HR, but not motion-based RR. Therefore, we tend to explore an intermediate compromised option, defocused camera, that reduces privacy risks (i.e. face cannot be recognized in such blurry images) but still allows processing of certain spatial contexts, e.g. motion analysis, posture estimation and in-bed detection.

To this end, we proposed a method to measure the cardio-respiratory parameters using defocused camera. It is possible to protect the privacy while monitoring physiological signals and certain visual contexts.
Keeping other camera parameters constant,
changing the focal length can blur the image, and a shorter focal length is preferred as it leads to a wider field of view, i.e. a longer focal length may not be able to capture the entire body. Therefore, with the fixed focus and aperture, we intentionally shortened the focal length to ensure that the camera cannot capture detailed facial information but still allows for skin-pixel analysis (see Fig.~\ref{fig:defocus}). 
We adopted a classical PPG-extraction framework to extract physiological signals from spatial-redundant pixels\cite{wang2014exploiting}, and a fully-automatic living-skin detection approach was incorporated to find the region of interest (RoI) for vital signs extraction~\cite{wang2017living}. The benchmark shows that HR measurement has a high correlation with the reference, RR measurement has a reasonable performance but can be improved further, i.e. by adjusting the illumination setup or changing the motion-based RR estimation to PPG-modulation based. 



\section{MATERIALS AND MEASUREMENT}
\subsection{Experiment}
The experiments in this study aim to:
\begin{itemize}
\item Validate the feasibility of using a defocused camera to measure HR and RR of the subjects while protecting their privacy.
\item Assess the impact of video blurring caused by lens defocus on the performance of vital signals monitoring.
\item Investigate the impact factors such as bedsheet occlusions and sleep postures on the performance of vital signals monitoring based on a defocused camera.
\end{itemize}

We simulated the night monitoring condition in a dark chamber, and a monochrome IR camera (IDS UI-3860CP-M-GL R2 with Sony's 2.1 MP sensor IMX290) and IR light source (850\,nm) were installed approximately 2 meters above the bed to provide a perspective view for capturing the entire torso of the sleeping subject.
To mimic real sleep scenarios, volunteers were instructed to sleep in the bed with three typical postures (left side, right side, and supine), and each posture lasted for 2 minutes. 
We defocused the camera to ensure that the images are sufficiently blurry (not possible for identification), and used a bedsheet to cover the subject to mimic the occlusion challenge. 
The infrared videos were recorded in four experimental conditions including clear-uncovered, clear-covered, blurry-uncovered, and blurry-covered.
All videos were recorded in an uncompressed data format, with the resolution of $548\times 968$ pixels at 20 frames per second. To benchmark the camera-based contactless vital signs monitoring, the reference physiological signals were recorded by Benevision N17 patient monitor (Mindray, China), which has three-lead ECG, PPG, and respiration.
Nineteen healthy volunteers (11 males and 8 females, average age: $22.7\pm 2.08$, average height: $168.0\pm 7.82$\,cm, average weight: $57.6\pm 10.38$\,kg) participated in this study. 
The study was approved by the Institutional Review Board of Southern University of Science and Technology, and written informed consent were obtained from the test subjects.

\begin{figure*}[t!]
        \centerline{\includegraphics[width=1.05\textwidth]
                {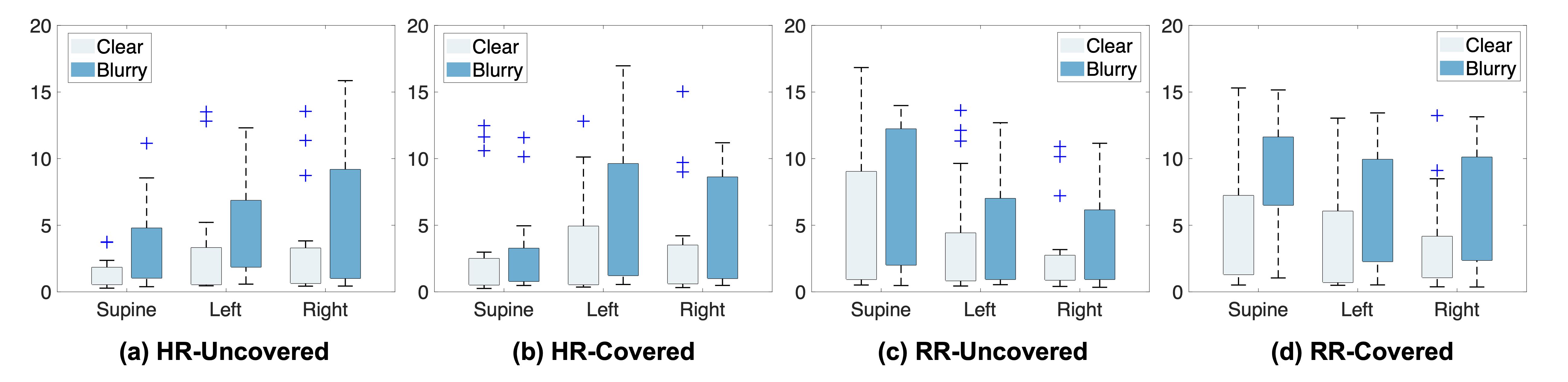}}
                \vspace{-0.2cm}
        \caption{ The MAE (bpm) of HR and RR between camera-based monitoring and reference under multiple conditions. 
        All boxplots show 6 MAE in 3 sleep postures with clear and blurry video conditions. (a) and (c) shows the MAE without bedsheet cover; (b) and (d) shows the MAE with bedsheet cover. 
        }
        \label{fig:boxes}
        \vspace{-0.2cm}
\end{figure*}

\subsection{Methods}
A classical PPG-extraction framework (see Fig.~\ref{fig:method}) known as Spatial Redundancy\cite{wang2014exploiting} was utilized in combination with the living-skin detection \cite{wang2017living} for vital signs extraction (e.g. HR and RR).
First, multi-scale block-based segmentation was used to partition the input video into image blocks with different sizes.
Second, for each image block, the PixFlow algorithm\cite{wang2022algorithmic} was used to derive the motion signals of the current image relative to the previous image in the vertical and horizontal directions, denoted as \textbf{\emph{dx}} and \textbf{\emph{dy}} respectively. The arctangent value of motion signals in both directions is calculated for subsequent motion analysis. Besides, the single IR-channel based PPG trace is generated by averaging the spatial pixels per image block and concatenating them in time, denoted as \textbf{\emph{C}}.
The motion signal and PPG signal are both segmented into multiple sliding windows for Fourier transform, and their power spectrum are generated for RR and HR calculation.
To identify the image blocks with high-quality signals, the Signal-to-Noise Ratio (SNR) was calculated to render heatmaps of motion signal and PPG signal. The breath energy is concentrated on the chest or abdomen, while the pulsatile energy dominates the facial skin. Thus, in the third step of automatic RoI selection, the SNR heatmaps are generated for PPG and breath separately (Fig.~\ref{fig:method}). Finally, the local signals selected from image blocks are combined into global signals for deriving HR and RR.

\subsection{Evaluation Metrics}
To validate the feasibility of the proposed method, we used the mean absolute error (MAE) to evaluate HR and RR measured from 19 adult subjects in the combination of multiple conditions (i.e. three sleep postures, with or without bedsheet occlusion, focused or de-focused camera setting). 
Since there are posture changes during the recording (i.e. for validating different sleep postures), significant body movements need to be addressed for continuous measurement. We use the intensity of motion signals as a metric to remove video segments with large movements for statistical evaluation.
The motion metric is computed by calculating the temporal standard deviation of motion signals. The subtle movements (including respiratory motion) during sleep typically have a motion intensity smaller than 1 in our method, thus a threshold = 1 is used to exclude large motions that are considered not relevant to sleep.
\begin{table}[]
        \caption{Mae of camera-based HR and RR monitoring in clear and blurry conditions}
        \renewcommand{\arraystretch}{1.3} 
        \label{table:data}
\centering
\begin{threeparttable} 
\begin{tabular}{c|cc|cc}
\hline \hline
\multirow{2}{*}{Subject} & \multicolumn{2}{c|}{\begin{tabular}[c]{@{}c@{}}Heart Rate\\  MAE ± SD (bpm)\end{tabular}} & \multicolumn{2}{c}{\begin{tabular}[c]{@{}c@{}}Respiratory Rate \\ MAE ± SD (bpm)\end{tabular}} \\ \cline{2-5} 
                         & \multicolumn{1}{c|}{Clear}                          & Blurry                            & \multicolumn{1}{c|}{Clear}                             & Blurry                              \\ \hline
1                        & 1.3±3.5                                             & 3.1±4.7                           & 13.0±13.1                                              & 6.1±4.7                             \\
2                        & 0.5±5.3                                             & 1.9±9.2                           & 1.4±1.9                                                & 5.3±4.6                             \\
3                        & 1.0±5.6                                             & 2.5±4.3                           & 1.2±1.3                                                & 6.4±5.2                             \\
4                        & 0.4±10.3                                            & 1.3±3.1                           & 0.6±0.7                                                & 1.6±1.6                             \\
5                        & 6.4±3.0                                             & 6.8±8.1                           & 0.9±1.3                                                & 3.4±3.8                             \\
6                        & 1.8±9.5                                             & 2.5±4.1                           & 1.4±1.8                                                & 1.7±2.4                             \\
7                        & 1.6±8.6                                             & 0.9±1.8                           & 6.3±4.8                                                & 9.8±5.5                             \\
8                        & 5.3±6.9                                             & 3.7±5.9                           & 3.3±3.6                                                & 7.5±6.3                             \\
9                        & 1.0±1.8                                             & 4.8±7.0                           & 1.0±1.1                                                & 1.3±1.4                             \\
10                       & 0.6±5.7                                             & 1.2±2.3                           & 1.1±1.8                                                & 11.7±5.9                            \\
11                       & 2.5±1.9                                             & 10.3±8.3                          & 6.0±3.8                                                & 10.9±6.4                            \\
12                       & 3.9±3.7                                             & 8.4±9.3                           & 1.4±2.3                                                & 4.9±4.1                             \\
13                       & 0.7±2.6                                             & 4.8±4.7                           & 1.0±1.3                                                & 0.9±1.4                             \\
14                       & 0.8±8.7                                             & 3.2±4.4                           & 2.5±3.4                                                & 6.1±5.2                             \\
15                       & 1.4±8.1                                             & 2.9±4.4                           & 3.7±3.5                                                & 6.6±4.2                             \\
16                       & 8.4±6.1                                             & 2.9±3.4                           & 13.6±6.3                                               & 5.6±5.0                             \\
17                       & 5.3±3.6                                             & 4.8±4.9                           & 9.4±6.6                                                & 2.0±2.1                             \\
18                       & 5.1±1.5                                             & 6.0±6.0                           & 5.7±4.9                                                & 9.8±5.4                             \\
19                       & 1.6±5.9                                             & 4.0±4.5                           & 2.4±3.6                                                & 9.3±6.2                             \\ \hline 
Overall                  & 2.7                                                 & 4.4                               & 4.0                                                    & 5.9                                 \\ \hline \hline
\end{tabular}
      \begin{tablenotes} 
		\item *“Clear” and “Blurry”: focused and de-focused condition.
            \item * MAE: mean absolute error; SD: standard deviation.
     \end{tablenotes} 
\end{threeparttable} 
        \vspace{-1.0cm}
\end{table}

\section{RESULTS AND DISCUSSION}
Table \ref{table:data} shows the MAE between the results of camera-based monitoring and the reference obtained from 19 subjects in clear and blurry videos.
Typically, the MAE of a reliable measurement of HR and RR is required to be less than 5\,bpm relative to the reference\cite{downey2019reliability}. The MAE of our measurements fluctuate around this standard.
The average MAE obtained in clear videos (by focused camera) is 2.7 bpm for HR, 4.0 bpm for BR, which indicates a good performance for camera-based monitoring in the focused condition. 
For the privacy protection, we extracted HR and RR from blurry videos (by de-focused camera) without detailed facial features.
The average MAE obtained from blurry videos is 4.4 bpm for HR, 5.9 bpm for BR, increased by 1.7 bpm and 1.9 bpm as compared to measurements from clear videos, respectively.
The MAE of HR and RR in blurry videos (see Table \ref{table:data} and Fig.~\ref{fig:boxes}) are generally increased due to the image blur. This is in line with our expectation as blurry images will create a mixture of skin and non-skin pixels that pollutes PPG measurement, and it also reduces sensitivity of pixel motion analysis.
As can be seen, the increase of MAE is larger in RR measurement than HR measurement, which suggests that the blurry effect has more negative impact on motion analysis than optical analysis of skin pulsatility. We also mention that since the used NIR-LED is in a low-power mode (resulting in lower image DC), the general performance of RR measurement is worse than HR measurement. 

Fig.\ref{fig:boxes} shows the MAE of HR and RR between camera-based monitoring and reference under various conditions. 
For HR monitoring, when subject is in the supine posture, MAE is the lowest in both clear and blurry conditions. 
This is because the camera can capture a larger area of facial skin in the supine posture.
However, for RR monitoring, no matter whether the video is clear or blurry, the MAE is the highest in the supine posture, which probably because that the respiratory motion is perpendicular to the camera sensor, leading to a weaker motion projected on the imaging plan orthogonal to the camera view.
The MAE of HR and RR under either the covered or uncovered condition has no significant difference, indicating the robustness to bedsheet occlusions in our setup.

\begin{figure}[t!]
        \centerline{\includegraphics[width=0.53\textwidth]
                {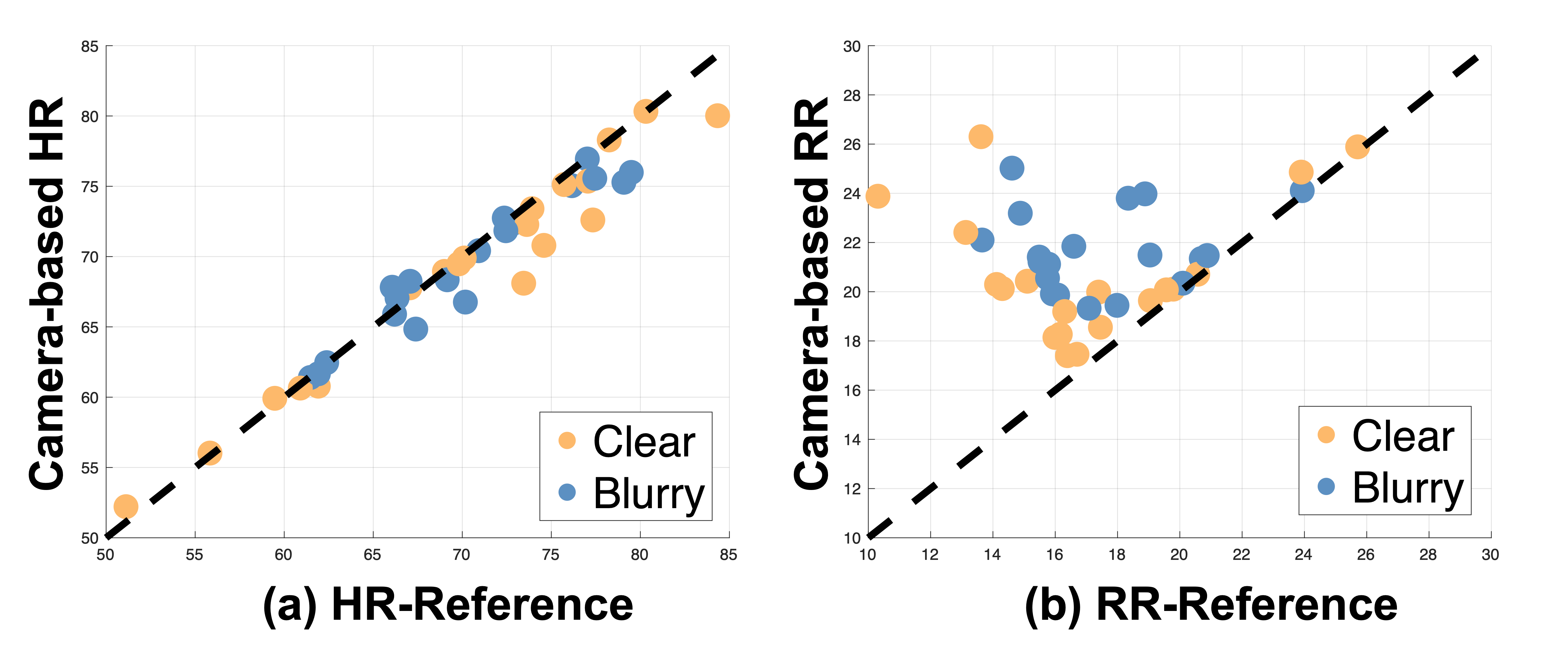}}
                \vspace{-0.1cm}
        \caption{Correlation between the camera measured HR and RR and the reference.} 
        \label{fig:hrrr}
        \vspace{-0.4cm}
\end{figure}

For HR measurement, although Table \ref{table:data} and Fig.~\ref{fig:boxes} indicate that the defocused camera somehow deteriorates the performance of camera-based monitoring, the measurements are still strongly correlated with reference (see Fig.~\ref{fig:hrrr}). 
This shows the feasibility of measuring HR in blurry videos without detailed facial information recorded by a defocused camera, which can physically protect a subjects' privacy during sleep.
Camera-based RR measurements are slightly higher than the reference in both clear and blurry conditions (see Fig.~\ref{fig:hrrr}). We suspect that this is due to the challenges of illumination in our setup since the MAE of RR estimation in both the clear and blurry conditions are relatively large. We calculated the Pearson correction of RR between these two conditions and found that they are strongly correlated ($P \leq 0.001$ and $R = 0.75$). It is possible to improve RR measurement by adjusting the setup or use other features (e.g. inter-beat interval) for RR extraction in the next step.

In summary, as compared with the focused camera, the increase of MAE in the de-focused condition indicates that image blur somehow deteriorates the vital signs monitoring, which is somehow expected. But the deterioration is still in a reasonable range. Our method seems to be robust to occlusion of bedsheet in both the focused and de-focused conditions. Although the performance of monitoring HR and RR using a defocused camera is inferior to that of a focused camera, the HR measurement of defocused camera has strong correlation with the reference, and RR measurement is in an acceptable range that can be improved further. We suggest to further explore the defocused camera for sleep monitoring considering the privacy protection. No sophisticated hardware modifications is required to adapt the setup except a lens that allows adjustment of the focal plane. In the next step, we will improve the RR measurement by adjusting the light source and using the respiratory modulations in PPG signal in addition to the pixel motions. We also tend to investigate other vital signs and contextual signals (e.g. sleep postures and activity) measured by the defocused camera, and conduct studies in real application scenarios involving sleep monitoring like sleep centers and senior centers.

\section{CONCLUSIONS}
To remotely monitor vital signs like HR and RR during sleep while protecting the subject privacy, we propose to use defocused cameras without complicated hardware modifications except the adjustment of the camera's focal length.
We applied spatial redundancy framework with living-skin detection to estimate HR and RR from blurry videos captured by a defocused camera. For blur caused by lens defocus, HR and RR monitoring are still in an acceptable range, especially the HR measured in blurry conditions still strongly correlates with the reference. The results shows that bedsheet occlusions do not have a major impact on the measurement performance. For sleep monitoring, defcoused camera can be considered for protecting the privacy while still giving reasonable results for vital signs monitoring.

\end{document}